\documentclass[twocolumn,epjc3]{svjour3} 
\RequirePackage[T1]{fontenc}
\smartqed
\RequirePackage{graphicx}
\RequirePackage{mathptmx}
\RequirePackage{flushend}
\RequirePackage[numbers,sort&compress]{natbib}
\RequirePackage[colorlinks,citecolor=blue,urlcolor=blue,linkcolor=blue]{hyperref}
\usepackage{amsmath}
\usepackage{widetext}
\usepackage{subfig}
\usepackage{float}
\usepackage{multirow}
\usepackage{balance}
\journalname{Eur. Phys. J. C}
\begin{document}

\title{Quasinormal modes of a charged spherical black hole with scalar hair for scalar and Dirac perturbations}
\author{Avijit Chowdhury\thanksref{e1,addr1}
        \and
        Narayan Banerjee\thanksref{e2,addr1}
}
\thankstext{e1}{e-mail: ac13ip001@iiserkol.ac.in}
\thankstext{e2}{e-mail: narayan@iiserkol.ac.in}
\institute{Department of Physical Sciences, Indian Institute of Science Education and Research Kolkata, Mohanpur Campus, Nadia, West Bengal 741246, India \label{addr1}}
\date{Received: date / Accepted: date}
\maketitle

\begin{abstract}
The quasinormal modes of charged and uncharged massive scalar fields and also of charged Dirac fields against the background of a charged spherical black hole endowed with a scalar hair have been investigated. Special emphasis has been given to the case where  negative scalar charge dominates over the electric charge of the black hole which mimics an Einstein-Rosen bridge. Except for the complete monotonic behaviour of the damping (imaginary part of the quasinormal mode) against the charge of the black hole as opposed to the existence of a peak for the pure RN case, the qualitative behaviour does not appreciably  change due to the presence of scalar hair.

\end{abstract}

\section{Introduction}
After the dream of detecting gravity waves came true, and that too from a merger of two black holes~\cite{ligo_PRL_2016}, the importance of a thorough investigation of the quasinormal modes in connection with the black hole perturbations cannot perhaps be exaggerated. These investigations started a long way back, through the work of Regge and Wheeler~\cite{regge_PR_1957} and  Vishveshwara~\cite{vishveshwara_PRD_1970, vishveshwara_Nature_1970}. The response of a black hole to a perturbation of an external field or the perturbation of the metric is manifested in the form of a damped wave emitted by the black hole, characterized by a complex frequency, called the quasinormal frequency. The real part of the frequency corresponds to the actual frequency of the wave motion while the imaginary part takes care of the damping factor. For excellent reviews, we refer to the works of Nollert~\cite{nollert_CQG_1999}, Kokkotas and Schmidt~\cite{kokkotas_LRR_1999} and Konoplya and Zhidenko~\cite{konoplya_RMP_2011}. 

Quasinormal modes (QNM) for a Schwarzchild black hole has been studied by Vishveshwara~\cite{vishveshwara_Nature_1970} and also by Davis, Ruffini, Press and Price~\cite{davis_PRL_1971}. QNMs for a Reissner-Nordstr\"{o}m black hole was first investigated by Gunter~\cite{gunter_PTRSLA_1980}. Investigations regarding QNMs for various kind of black holes are already there in the literature. Dreyer discussed the QNMs, area spectrum and entropy of a black hole and also fixed the value of the Immirizi parameter which arises in Loop quantum gravity~\cite{dreyer_PRL_2003}. Cardoso and Lemos discussed the QNMs of a BTZ black hole~\cite{cardoso_PRD_2001_1} and also Schwarzchild-AdS black holes~\cite{cardoso_PRD_2001_2}. The latter had been discussed by Horowitz and Hubeny~\cite{horowitz_PRD_2000} also. QNMs for a  near extremal black hole has been investgated by Starinets~\cite{starinets_PRD_2002} and by Cardoso and Lemos~\cite{cardoso_PRD_2003}. QNMs for a Gauss-Bonnet black hole has been discussed by Chakrabarti~\cite{sayan_GRG_2007}.

The purpose of the present work is to investigate the QNMs of a black hole endowed with a scalar hair. We pick up the example which has been very recently given by Astorino~\cite{astorino_prd_2013}. The black hole has both electric charge and scalar charge. The scalar field part is basically the same one as that given by Bekenstein~\cite{bekenstein_AP_1974, bekenstein_AP_1975}, but in a much more useful form. The metric is qualitatively same as the Reissner-Nordstr\"{o}m (RN) metric. The ``scalar charge'' comes only as an additive correction to the electric charge, so nothing new comes out of it as such to start with. But the scalar charge comes with a power unity in the metric as opposed to the quadratic appearence of the electric charge.  Thus one can set the scalar charge $s<-e^2$ so as to get the $\frac{1}{r^2}$ term with a negative coefficient. This is a very simple realization of a ``mutated Reissner-Nordtsr\"{o}m'' metric leading to the Einstein-Rosen bridge~\cite{rosen_PR_1935} or a so called wormhole. The present work deals with such a metric, with primarily negative values of $s$. The perturbation of massless and massive uncharged/charged scalar particles and massless charged Dirac particles and the QNMs generated by the perturbations in the vicinity of a mutated Reissner-Nordstr\"{o}m black hole are discussed in the present work. The continued fraction method (see Refs.~\cite{leaver_PRSLA_1985,leaver_PRD_1990,leaver_PRD_1991}) has been adopted. The fundamental mode is the dominating one in the signal and only that mode is dealt with.

In almost all the cases both the frequency and the damping rate decrease with the magnitude of the negative scalar charge. For massive scalar field, the damping rate falls off sharply compared to the massless case whereas the real frequency falls off at a much slower rate. For charged fields, the oscillation frequency and the damping rate is more for higher values of the field charge.

The paper is organized in the following way. We start with a brief description of the background spacetime in section 2. In section 3, we briefly describe the continued fraction technique and discuss the QNMs for both uncharged and charged scalar fields close to a mutated RN black hole. Section 4 includes a discussion on the the QNMs of massless charged Dirac field around an RN black hole endowed with a scalar hair. The fifth and final section contains a summary and discussion on the results obtained. As it is already mentioned, the work is done using the continued fraction method. It has also been worked out using a 3rd order WKB method (see Refs.~\cite{iyer_PRD_1987_1,iyer_PRD_1987_2}), but not mentioned in the text. To facilitate a comparison, we include a table showing the results of the two methods for one example as an appendix.

\section{Background Spacetime} \label{sec_2}
Starting from the action of general relativity coupled to a Maxwell field $F^{\mu \nu}$ and conformally coupled to a scalar field $\psi$,
\begin{equation}\label{eq_action}
\begin{split}
I=\frac{1}{16\pi G}\int d^4 x\sqrt{-g}\left[ R-F_{\mu \nu}F^{\mu \nu}\right.\\ \left.-8\pi G\left(\bigtriangledown_\mu \psi \bigtriangledown^\mu \psi+\frac{R}{6}\psi^2\right)\right]
\end{split}
\end{equation}
Astorino~\cite{astorino_prd_2013} arrived at the Reissner-Nordstr\"{o}m black hole of mass $M$ and charge $e$ endowed with a scalar hair $s$,
\begin{equation}\label{eq_metric}
ds^2=-f\left(r \right)dt^2+{f\left(r \right)}^{-1}dr^2+r^2 \left( d\theta^2+\sin^2{\theta} d\phi^2 \right),
\end{equation}
where 
\begin{eqnarray}
\label{eq_f(r)}
f\left( r \right)&=&\left(1-\frac{2M}{r}+\frac{e^2+s}{r^2}\right)\hspace*{0.2cm} \mbox{and}\\
\psi &=&\pm \sqrt{\frac{6}{8 \pi G}}\sqrt{\frac{s}{s+e^2}}.
\end{eqnarray}

The net stress-energy tensor looks like
\begin{equation}\label{eq_stressenergy}
T^\mu_\nu=\frac{e^2+s}{r^4} diag\left(-1,-1,1,1\right) . 
\end{equation}

It is interesting to note that the scalar field $\psi$ has a constant value. Still it gives a non-trivial contribution to the metric because of its nonminimal coupling with geometry in the form $\frac{R}{6}\psi^2$ in the action~(\ref{eq_action}). 
The scalar hair $s$ is a primary hair since the scalar field $\psi$ survives even in the absence of the electromagnetic field. It is easy to note from relation~(\ref{eq_stressenergy}) that the trace of the energy momentum tensor due to the scalar field alone is also zero. Thus the existence of this hair is completely consistent with the theorem given in Ref.~\cite{narayan_pramana_2015}. In the range $0>s>-e^2$, the scalar field has an imaginary value. In such a case, the kinetic part in the action should have been written as ${\nabla}_{\mu}{\psi}^{*}{\nabla}^{\mu} \psi$. However, it hardly matters in the present case as the kinetic part becomes trivial as $\psi$ is a constant. As discussed earlier, one of the principal motivation of the work is to look at the QNMs for a mutated Reissner-Nordstr\"{o}m black hole which requires $s<-e^2$, the question of a complex scalar field will not arise.

The present black hole given by the solution~(\ref{eq_metric}) and (\ref{eq_f(r)}), henceforth referred to as the Reissner-Nordstr\"{o}m-scalar hair (shRN) black hole, is also characterised by an inner Cauchy horizon $\left(r_{-} \right)$ and an outer event horizon $\left(r_{+} \right)$. The horizons of the shRN black hole are located at
\begin{eqnarray}
r_{+}&=&M+\sqrt{M^2-e^2-s},\\
r_{-}&=&M-\sqrt{M^2-e^2-s}.
\end{eqnarray}
The maximum value of the scalar (or electric) charge is determined by the extremality condition,
\begin{equation}\label{eq_extremality}
\sqrt{M^2-e^2-s} = 0.
\end{equation}
For $s=-e^2$ the shRN spacetime reduces to a Schwarzschild black hole with the event horizon at $r_{+}=2M$. The mutated RN spacetime is also characterised by a single event horizon as $r_{-}$ is negative and of no physical significance.
\section{Massive scalar field around a charged black hole with scalar hair}
In this section we discuss the dynamics of a massive charged scalar field in the background of an shRN black hole and study the fundamental $(n=0)$ mode of the  quasinormal spectrum of the field around the black hole.
\subsection{Field dynamics}
The dynamics of a massive charged test scalar field $\Phi$ of mass $\mu$ and electric charge $q$ in the background~(\ref{eq_metric}) is governed by the Klein-Gordon equation,
\begin{equation}\label{eq_KG}
[\left(\nabla^\nu-iqA^\nu\right)\left(\nabla_\nu-iqA_\nu\right)-\mu^2]\Phi=0
\end{equation}
where $A_\nu=-\delta^0_\nu e/r$ is the electromagnetic vector potential of the black hole.
We can decompose the field $\Phi$ as
\begin{equation}\label{eq_decom}
\Phi_{lm}\left(t,r,\theta,\phi\right)=e^{-i\omega t}S_{lm}\left(\theta\right)R_{lm}\left(r\right)e^{i m \phi},
\end{equation}
where $\omega$ is the conserved frequency, $l$ is the spherical harmonic index and $m$ ($-l\leq m\leq l$) is the azimuthal harmonic index. Hereafter we will drop the subscripts $l$ and $m$ for brevity.

With the decomposition~(\ref{eq_decom}) one can separate the Klein-Gordon equation~(\ref{eq_KG}) into a radial and an angular equation with the separation constant $K_l=l\left( l+1 \right)$. The radial Klein-Gordon equation is given by
\begin{equation}\label{eq_radialKG}
\frac{d}{dr}\left(\Delta \frac{dR}{dr}\right)+\frac{U}{\Delta} R=0,
\end{equation}
where $\Delta=r^2 f\left(r\right)$ and 
\begin{equation}
U=\left( \omega r^2-e q r \right)^2-\Delta \left[ \mu ^2 r^2+l \left( l+1 \right)\right].
\end{equation}
If we define a new radial function $\zeta=r R$ and adopt the tortoise coordinate $r_{*}$ $\left( \right.$defined by, $d r_{*}=dr/f\left( r \right)\left.\right)$, mapping the semi infinite region $\left[r_{+},\infty\right)$ into $(-\infty,\infty)$, then the radial Klein-Gordon equation~(\ref{eq_radialKG}) becomes
\begin{equation}\label{eq_trts}
\frac{d^2 \zeta}{d r_{*}^2}+W\left(\omega, r\right) \zeta=0,
\end{equation}
where
\begin{equation}
\begin{split}
W\left(\omega, r\right)&=\left(\omega -\frac{e q}{r}\right)^2\\
& -f(r) \left(-\frac{2 \left(e^2+s\right)}{r^4}+\frac{2 M}{r^3}+\frac{(l+1) l}{r^2}+\mu ^2\right).
\end{split}
\end{equation}
In the asymptotic limit equation~(\ref{eq_trts}) can be solved analytically with the quasinormal mode (QNM) boundary conditions of purely ingoing waves at the horizon $\left(r_{*}\rightarrow-\infty\right)$ and purely outgoing waves at spatial infinity $\left( r_{*}\rightarrow \infty \right)$,
\begin{equation}\label{eq_bc}
\zeta \approx 
\begin{cases}
 e^{-i \left(\omega-\frac{e q}{r_{+}} \right) r_*}&\mbox{ as \hspace*{2mm}}r_* \rightarrow -\infty \\
r_{*}^{-i e q} e^{i\Omega r_{*}}&\mbox{ as \hspace*{2mm}}r_* \rightarrow \infty,
\end{cases}
\end{equation}
where  $\Omega=\sqrt{\omega^2-\mu^2}$. \\Equation~(\ref{eq_trts}) together with the boundary conditions~(\ref{eq_bc}) becomes an eigenvalue problem with complex eigenvalues $\omega$ representing the quasinormal frequencies.
\subsection{Continued Fraction technique}
In 1985, Leaver~\cite{leaver_PRSLA_1985,leaver_PRD_1990,leaver_PRD_1991} inspired by a seminal work of Jaff\'{e}~\cite{Jaffe_Z_Phys_1934} on the calculation of the electronic spectra of hydrogen molecular ion, proposed a very accurate method for finding out the QNM frequencies of black holes.

To implement Leaver's method we start with equation~(\ref{eq_radialKG}) and observe that it has two regular singularities at $r_{+}$ and $r_{-}$ and an irregular singularity as $r\rightarrow \infty$ .

We can write a solution to equation~(\ref{eq_radialKG}) with the desired behaviour at the boundaries as
\begin{equation}\label{eq_scalar_ansatz}
R=e^{i \Omega r} (r-r_{-})^\rho\sum_{n=0}^{\infty}a_n u^{n+\delta},
\end{equation}
where $u=\frac{r-r_{+}}{r-r_{-}}$, $\rho=\frac{i \left(i \Omega+M \left(\Omega^2+\omega^2\right)-e q \omega\right)}{\Omega}$	and $\delta=-\frac{i r_{+}^2 \left(\omega -\frac{e q}{r_{+}}\right)}{r_{+}-r_{-}}$.\\
Substituting the ansatz~(\ref{eq_scalar_ansatz}) into equation~(\ref{eq_radialKG}) we arrive at the following three term recurrence relations, satisfied by the coefficient $a_n$
\begin{eqnarray}
\alpha_0 a_1 +\beta_0 a_0 & = & 0,\\
\alpha_n a_{n+1} +\beta_n a_n+\gamma_n a_{n-1} & = & 0,
\end{eqnarray}
where $\alpha_n$, $\beta_n$ and $\gamma_n$ are given by,

\begin{widetext}
\begin{align}
\alpha_n&=-\frac{\left(n+1\right)^2 r_{-}+(n+1)r_{+} \left(-2 i e q-n+2 i r_{+} \omega -1\right)}{r_{+}-r_{-}},\\
\beta_n&=
\begin{aligned}[t]
&\frac{1}{2 \Omega (r_{-}-r_{+})}\left[ r_{+} \left\lbrace 2 \left(-2 e^2 q^2 \left(\Omega+\omega \right)+i e (2 n+1) q \left(2 \Omega+\omega \right)+\left(l \left(l+1\right)+2 n^2+2 n+1\right) \Omega\right)\right.\right.\\
&\left.\left.+r_{+} \left(4 \omega  \left(\Omega+\omega \right) (3 e q-2 i n-i)+3 i \mu ^2 (2 i e q+2 n+1)\right)+2 r_{+}^2 \left(\mu ^2 \left(\Omega+3 \omega \right)-4 \omega ^2 \left(\Omega+\omega \right)\right)\right\rbrace \right.\\
&\left.-2 r_{-} \left\lbrace i r_{+} \left(-2 (2 n+1) \omega ^2+\mu ^2 (i e q+4 n+2)\right)+i e (2 n+1) q \omega +\left(l \left(l+1\right)+2 n^2+2 n+1\right) \Omega \right.\right.\\
&\left.\left.+\mu ^2 r_{+}^2 \left(\Omega+\omega \right)\right\rbrace +i \mu ^2 (2 n+1) r_{-}^2\right],
\end{aligned}\\[\jot]
\gamma_n&=
\begin{aligned}[t]
&\left[\frac{i \left\lbrace e q \omega -\frac{1}{2} \left(\Omega^2+\omega^2\right) \left(r_{-}+r_{+}\right)\right\rbrace }{\Omega}+i e q+n-i \omega  (r_{-}+r_{+})\right]\\
&\left[n-\frac{i \left\lbrace-2 \left(r_{+}-r_{-}\right) \left(e q \omega -\frac{1}{2}\left(r_{-}+r_{+}\right)\left(\Omega^2+\omega ^2\right)\right)+\Omega \left(r_{-}+r_{+} \right) \left(\omega  \left(r_{-}+r_{+}\right)-2 e q\right)+\omega  \Omega \left(r_{-}-r_{+}\right)^2\right\rbrace}{2 \Omega \left(r_{+}-r_{-}\right)}\right].
\end{aligned}
\end{align}
\end{widetext}
The convergence of the series~(\ref{eq_scalar_ansatz}) requires 
the recursion coefficients to satisfy an infinite continued fraction relation
\begin{equation}\label{eq_scalar_CF}
0=\beta_0-\frac{\alpha_0 \gamma_1}{\beta_1-} \frac{\alpha_1 \gamma_2}{\beta_2-}\cdots\frac{\alpha_n \gamma_{n+1}}{\beta_{n+1}-}\cdots
\end{equation}
The solution to this infinite continued fraction equation gives the QNM frequencies. The  continued fraction relation~(\ref{eq_scalar_CF}) can be inverted any number of times. Numerically, the $n^{th}$ QNM frequency is defined to be the most stable root of the $n^{th}$ inversion of the continued fraction relation,
\begin{equation}\label{eq_scalar_CF_inv}
\begin{split}
\beta_n-\frac{\alpha_{n-1}\gamma_{n}}{\beta_{n-1}-} \frac{\alpha_{n-2}\gamma_{n-1}}{\beta_{n-2}-}\cdots\frac{\alpha_{0}\gamma_{1}}{\beta_{0}}=\frac{\alpha_{n} \gamma_{n+1}}{\beta_{n+1}-} \frac{\alpha_{n+1} \gamma_{n+2}}{\beta_{n+2}-}\cdots, \\ \left(n=1,2,3,4\cdots \right).
\end{split}
\end{equation}
In practice the infinite continued fractions in equations~(\ref{eq_scalar_CF}, \ref{eq_scalar_CF_inv}) are truncated at some large truncation index, $N$. Nollert~\cite{Nollert_PRD_1993} has shown that the ``error'' due to truncation can be minimised and the convergence of the method can be improved by a wise choice of the ``remaining'' part of the infinite continued fraction , $R_N=-\frac{a_{N+1}}{a_N}$, which in turn satisfies the recurrence equation,
\begin{equation}\label{eq_Nollert}
R_N=\frac{\gamma_{N+1}}{\beta_{N+1}-\alpha_{N+1}R_{N+1}}.
\end{equation}
Assuming that $R_N$ can be expanded in a power series of $N^{-1/2}$,
\begin{equation}\label{eq_nollert_series}
R_N=\sum_{k=0}^{\infty}C_{k}~N^{-k/2},
\end{equation} 
we  obtain the first three coefficients $C_k$ as,\\
$C_0=-1$, $C_1=\sqrt{2i\left(r_{-}-r_{+}\right)\left(\omega^2-\mu^2\right)^{1/2}}$ and\\$C_2=-\frac{i \left(e q \omega -\mu ^2 M\right)}{\sqrt{\omega ^2-\mu ^2}}+2 i r_{+} \sqrt{\omega ^2-\mu ^2}+\frac{3}{4}$.

\subsection{Numerical Results}
We first study the fundamental QNMs due to uncharged massive scalar field then we  add electric charge to the perturbing field and study the effect of the scalar hair on the QNMs.
For the sake of numerical simplicity we scale the mass of the black hole to unity.
\subsubsection{Uncharged massive scalar field}
For an uncharged scalar field $(q=0)$ in the shRN background we assume, without any loss of generality, the constant electric charge of the black hole to be zero,$(e=0)$. The function $W$ appearing in equation~(\ref{eq_trts}) can now be written as $W(\omega,r)=\omega^2-V(r)$ with
\begin{equation}
V(r)=f(r) \left(-\frac{2 s}{r^4}+\frac{2 M}{r^3}+\frac{(l+1) l}{r^2}+\mu ^2\right).
\end{equation}

\begin{figure*}
\centering
\includegraphics[width=0.99\linewidth,height=0.3\linewidth]{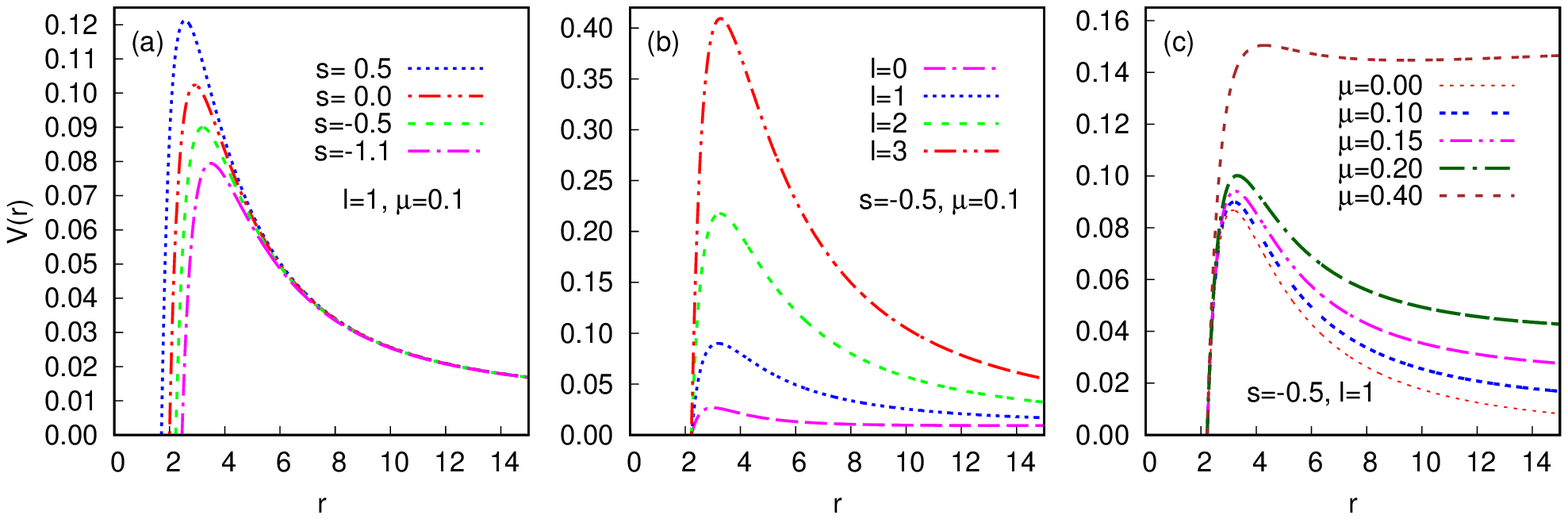}
\caption{Panel (a) shows the variation of $V(r)$ with $r$ for $l=1$ and $\mu=0.1$ for different values of $s$ as indicated. Panels (b) and (c) shows the variation of $V(r)$ with $r$ for $s=-0.5$, $\mu=0.1$ and for $s=-0.5$, $l=1$, respectively. Each curve in  (b) corresponds to a particular value of $l$ and each curve in (c) corresponds to a particular value of $\mu$ as indicated.}
\label{fig_1}
\vspace*{0.5cm}
\includegraphics[width=1\textwidth]{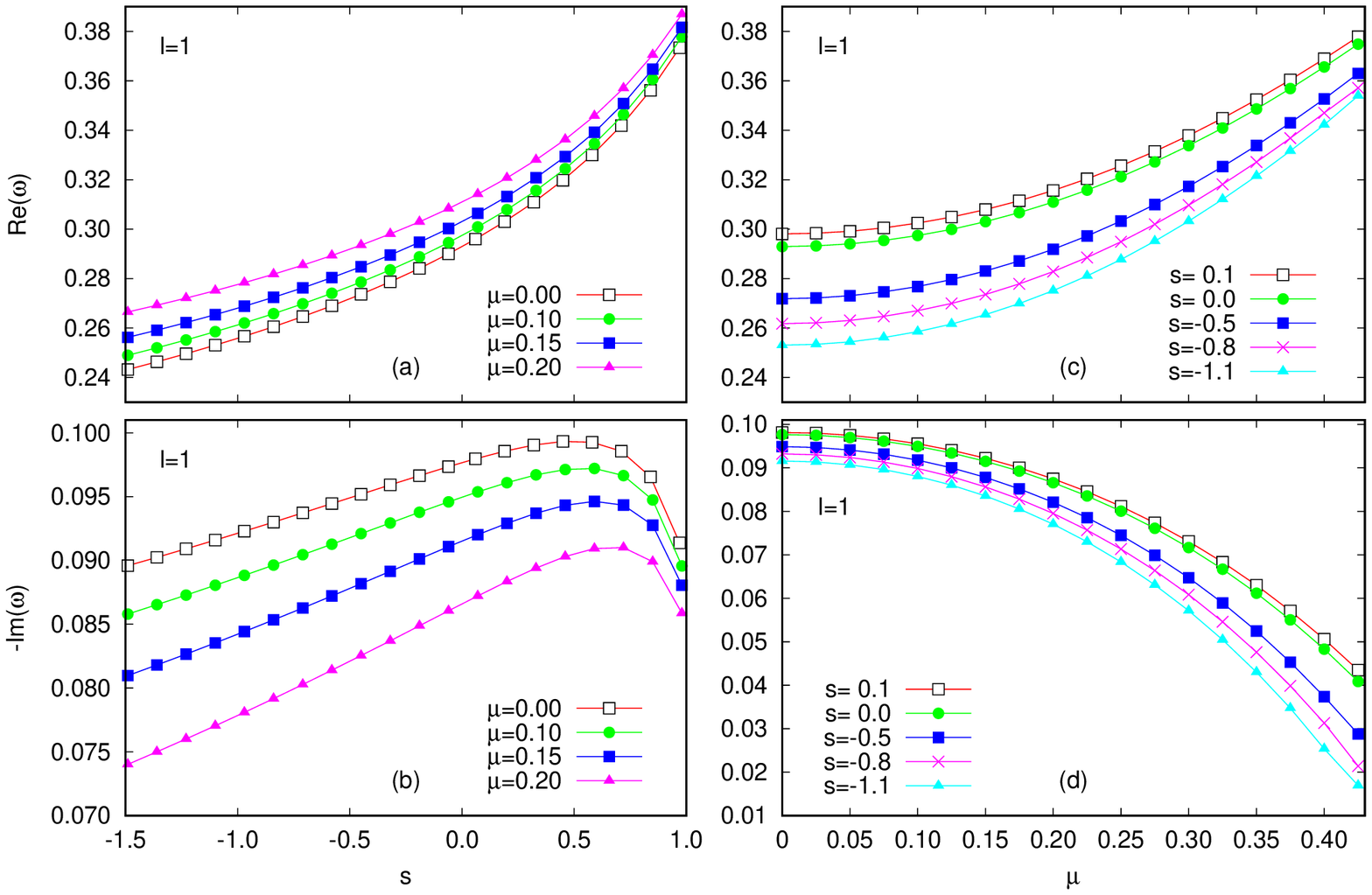}
\caption{Panels (a) and (b), respectively, show the real and imaginary parts of the fundamental (scalar) QN frequency as a function of $s$ for $l=1$ with each curve corresponding to a particular value of $\mu$ as indicated. Panels (c) and (d), respectively, show the real and imaginary parts of fundamental (scalar) QN frequency as a function of $\mu$ for $l=1$ with each curve corresponding to a particular value of s as indicated. The curve with $s=0.1$ corresponds to an RN black hole with $e\simeq 0.316$, while the curve with $s=0$ corresponds to a Schwarzschild black hole.}
\label{fig_2}
\end{figure*}

\begin{figure*}
\centering
\includegraphics[width=1\linewidth]{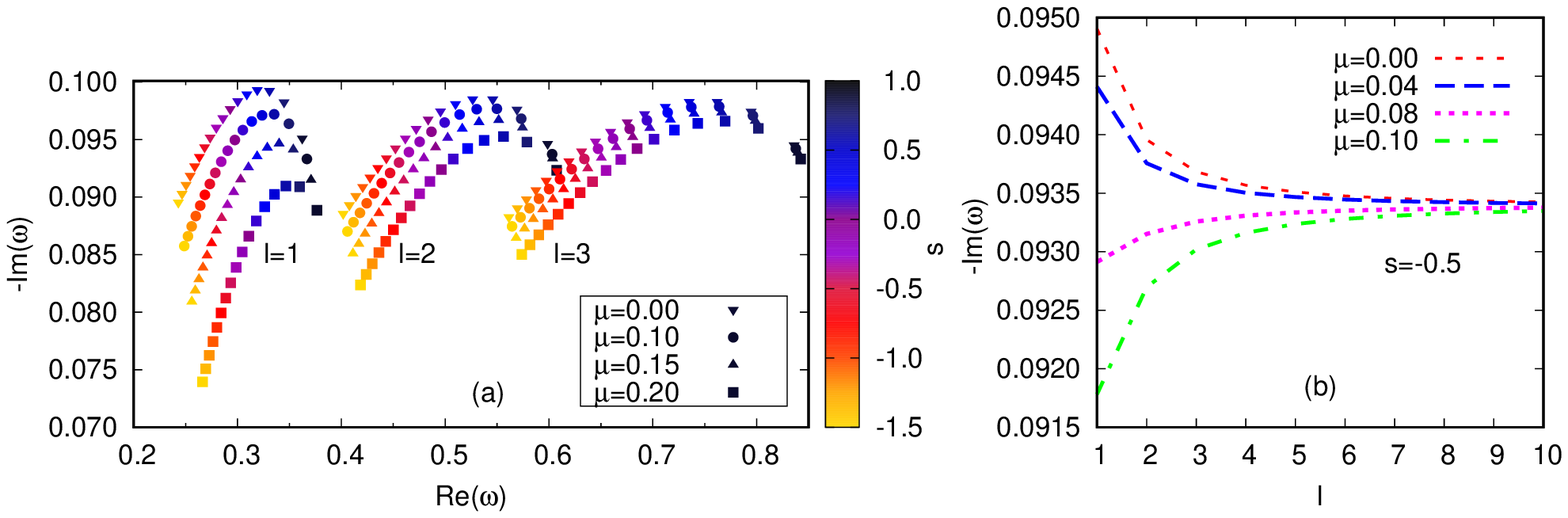}
\caption{Panel (a) shows the variation of the imaginary part of the fundamental (scalar) QN frequency with the Real part for different values of $s$ and $l$. For a given $l$, each curve corresponds to a particular value of $\mu$ as indicated. Panel (b) shows the imaginary part of the fundamental (scalar) QN frequency as a function of $l$ with $s=-0.5$ for different values of $\mu$ as indicated.}
\label{fig_3}
\vspace*{0.5cm}
\includegraphics[width=1\textwidth]{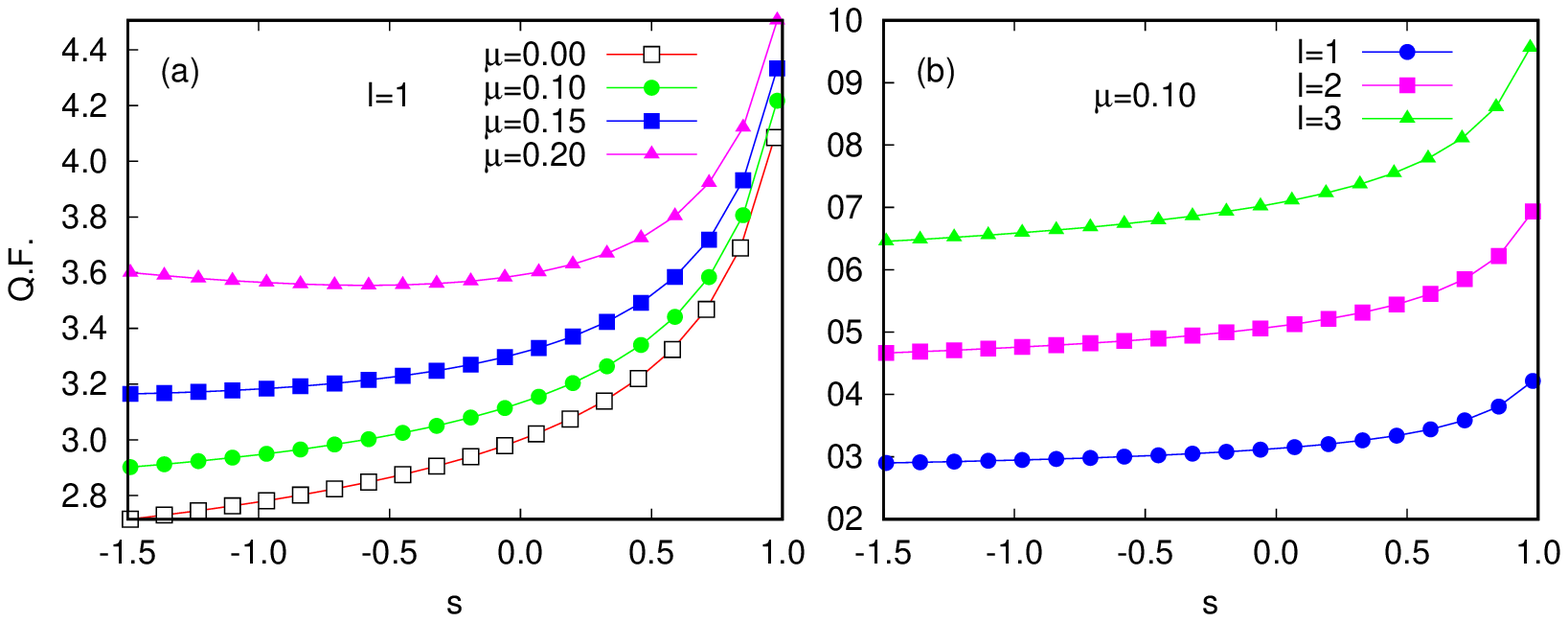}
\caption{Panels (a) and (b) show the quality factor as a function of $s$ with $l=1$ for different values of $\mu$ and with $\mu=0.1$ for different values of $l$, respectively. }
\label{fig_4}
\end{figure*}

The potential of the of the shRN black hole for different values of the scalar charge, field mass and multipole index are shown in Fig.~\ref{fig_1}.

In Figs.~\hyperref[fig_2]{\ref*{fig_2}(a)} and \hyperref[fig_2]{(b)}, we show the behaviour of the real and imaginary parts of the fundamental QN frequency with the scalar charge for a particular multipole index $(l=1)$ and different field masses. We observe that for $s<0$, the magnitude of both the real and imaginary parts of the  QN frequency decrease with the absolute value of the scalar charge. This implies that the real oscillation frequency as well as the damping rate decrease with increasing magnitude of the negative scalar charge. For $s>0$, the spacetime~(\ref{eq_metric}) effectively behaves as an RN black hole of unit mass and electric charge, $e=\sqrt{s}$, showing a distinct peak in the magnitude of the imaginary part of the fundamental quasinormal frequency (see Refs.~\cite{konoplya_PRD_2002,konoplya_PLB_2002}). We also observe that for a particular value of the scalar charge, the real part of the QN frequency increases with the field mass whereas the magnitude of the imaginary part decreases. This behaviour is manifested more clearly in Figs.~\hyperref[fig_2]{\ref*{fig_2}(c)} and \hyperref[fig_2]{(d)}, where we observe that for sufficiently large field masses, the imaginary part of the QN  frequency becomes vanishingly small. This results in long lived, purely real modes in the quasinormal spectrum, called \textit{quasi-resonance modes}~\cite{ohashi_cqg_2004}.
We also note that as the scalar charge changes from positive to negative,\textit{quasi-resonance} occurs at lower field masses with smaller real frequencies.

Fig.~\hyperref[fig_3]{\ref*{fig_3}(a)} shows a compact view of the behaviour of the real and imaginary parts of the QN frequency with the scalar charge for different values of the multipole number and field mass. We note that, as the multipole number increases the real part of the fundamental QN frequency increases  and so does the imaginary part, but only for higher field masses. This behaviour of the imaginary part of the QN frequency can be seen more clearly in Fig.~\hyperref[fig_3]{\ref*{fig_3}(b)} where we note that for lower field masses, the damping rate decreases with the multipole index whereas for higher field masses, it increases with the multipole index. For large values of the multipole number, the damping rate is almost insensitive to the field mass.

Following Ref.~\cite{sayan_EPJC_2009} we define the \textit{Quality Factor} as \\ $Q.F. \sim \vert \frac{\omega_{Re}}{\omega_{Im}}\vert$. In Fig.~\hyperref[fig_4]{\ref*{fig_4}(a)}, we observe that for a given multipole index and for large positive values of the scalar charge the quality factor decreases sharply, however for smaller values of the scalar charge it decreases very gradually. For massless field, the gradual decrease of the quality factor continues to persist for negative values of the scalar charge as well. However, beyond a certain value of the field mass $\mu$, it tends to increase for high negative values of $s$, the plot corresponding to $\mu=0.2$ in Fig.~\hyperref[fig_4]{\ref*{fig_4}(a)} reveals this feature. The quality factor has higher values for higher multipole indices (see Fig.~\hyperref[fig_4]{\ref*{fig_4}(b)}). The quality factor is a measure of the product of the frequency and the ring down time of a black hole radiation, and is an important tool to figure out the black hole parameters~\cite{cardoso_thesis}.

\subsubsection{Charged scalar field}
\begin{figure*}
\centering
\includegraphics[width=0.99\textwidth]{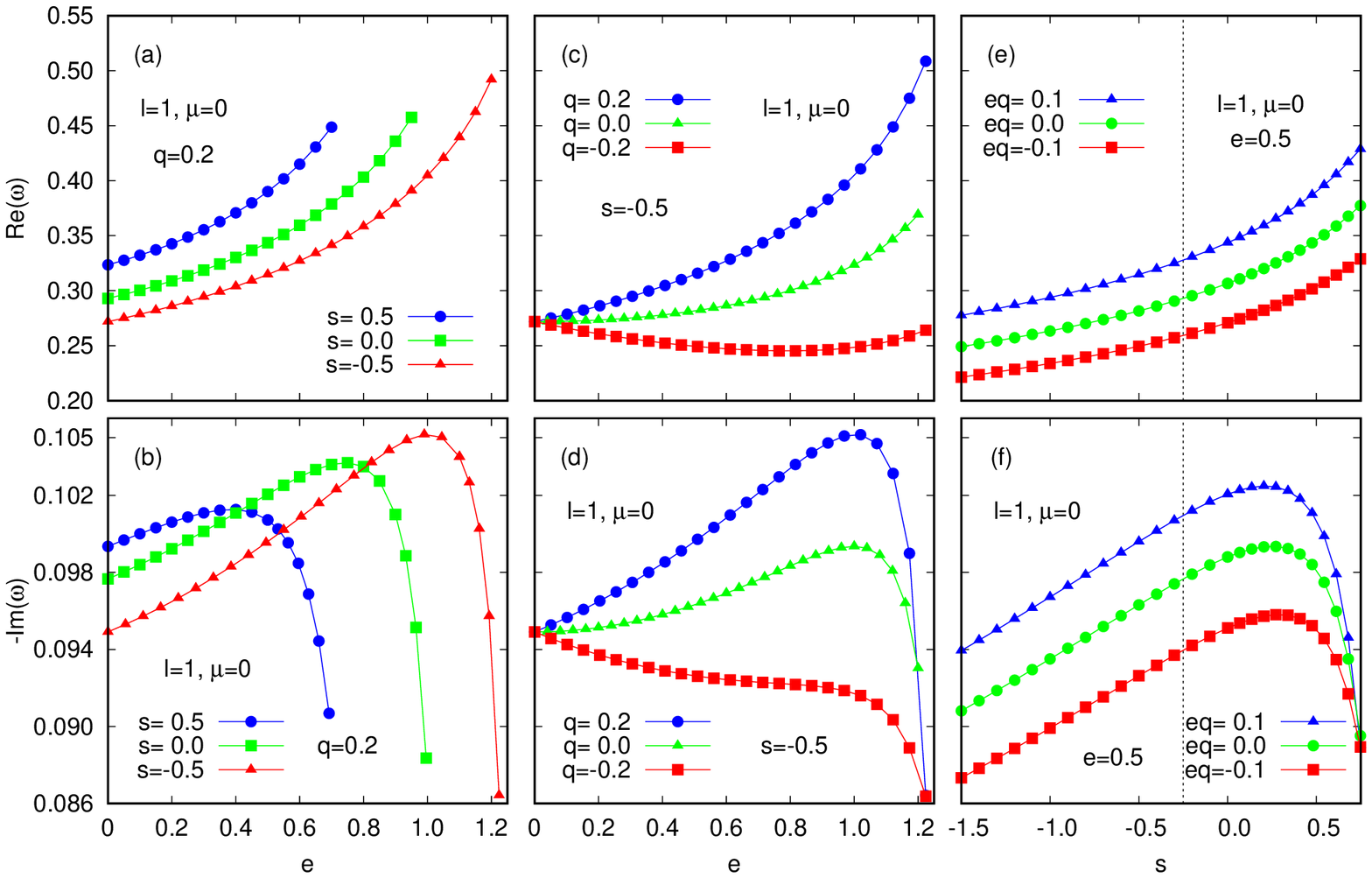}
\caption{Panels (a) and (b), respectively, show the real and imaginary parts of the fundamental (scalar) QN frequency as a function of $e$ for $l=1$, $\mu=0$ and $q=0.2$ with each curve corresponding to a particular value of $s$ as indicated. Panels (c) and (d), respectively, show the real and imaginary parts of the fundamental (scalar) QN frequency as a function of $e$ for $l=1$, $\mu=0$ and $s=-0.5$ with each curve corresponding to a particular value of $q$ as indicated. Panels (e) and (f), respectively, show the real and imaginary parts of the fundamental (scalar) QN frequency as a function of $s$ for $l=1$, $\mu=0$ and $e=0.5$ wih each curve corresponding to a particular value of $eq$ as indicated. The vertical line denotes the value of $s$ (=-0.25) below which the spacetime behaves as mutated RN.}
\label{fig_5}
\vspace*{0.5cm}
\includegraphics[width=1\textwidth]{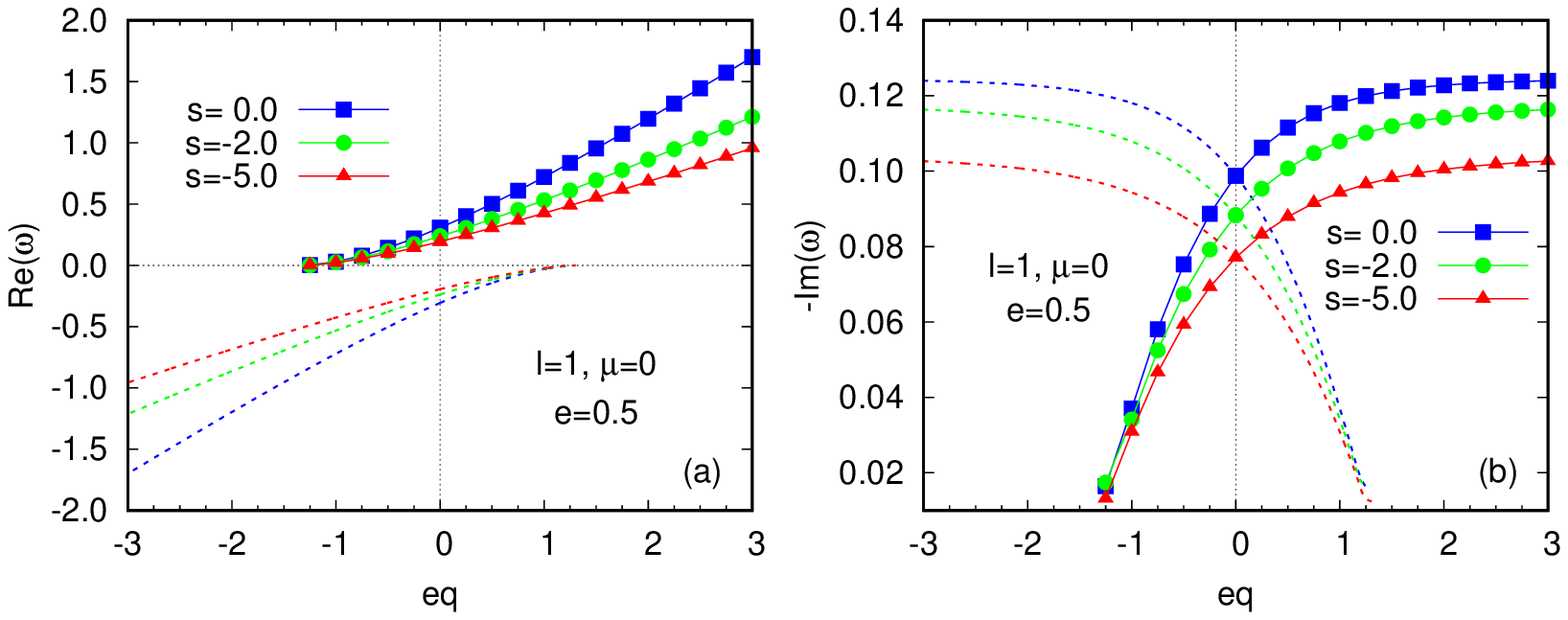}
\caption{Panels (a) and (b), respectively, show the real and imaginary parts of the fundamental (scalar) QN frequency as a function of $eq$ for $l=1$, $\mu=0$ and $e=0.5$. Each curve in each panel corresponds to a particular value of $s$ as indicated. The dashed lines represent the symmetric curves due to the simultaneous transformation $eq\rightarrow -eq$ and $\omega\rightarrow -\omega^{*}$.}
\label{fig_6}
\end{figure*}
The presence of the scalar hair changes the frequency and damping rate of the QN spectrum of  charged scalar fields as well. In Figs.~\hyperref[fig_5]{\ref*{fig_5}(a)} and \hyperref[fig_5]{(b)}, we observe that compared to the RN black hole, for fixed non-zero values of $e$ and $q$, the magnitude of both the real and imaginary parts of the fundamental QN frequency are higher for positive values of the scalar charge and lower for negative values.

Konoplya~\cite{konoplya_PRD_2002} observed that for an RN black hole, the imaginary part of the QN frequency, for any given value of the field charge, approaches that for the uncharged field as the extremal limit is approached. Apart from a similar observation in the presence of a scalar hair (see Figs.~\hyperref[fig_5]{\ref*{fig_5}(c)} and \hyperref[fig_5]{(d)}) as well, we note from Fig.~\hyperref[fig_5]{\ref*{fig_5}(f)} that such convergence of the imaginary part of the fundamental QN frequency occurs for any given value of the black hole electric charge, as the  maximal value of the scalar charge is approached.  This maximal value is determined by the extremality condition~(\ref{eq_extremality}). We further observe that, for $s>-e^2$, the magnitude of the imaginary part of the fundamental QN frequency shows a distinct peak whereas for $s<-e^2$, it decreases monotonically with the magnitude of $s$. The corresponding behaviour of the real part of the QN frequency with the scalar charge is shown in Fig.~\hyperref[fig_5]{\ref*{fig_5}(e)}.

The symmetry of the QNMs with respect to the transformation $\left(eq\rightarrow -eq, \omega\rightarrow -\omega^{*}\right)$ is depicted in Fig.~\ref{fig_6}.  Fig.~\hyperref[fig_6]{\ref*{fig_6}(a)} also highlights the existence of a critical value of $|eq|$ at which the real part of the QN frequency vanishes. However, such a behaviour of the QN frequency is not new and has been previously observed for the RN black hole for charged scalar and Dirac fields (see Refs.~\cite{konoplya_PRD_2013,richartz_PRD_2014}). We note in particular, that the critical value of $|eq|$ is almost unaffected by the presence of the scalar hair and does not change with the the black hole electric charge. For an RN black hole with unit multipole index, the critical value is $|eq|\approx1.3$.
\section{Charged Dirac field around a charged black hole with scalar hair}
The dynamics of a massless charged Dirac field propagating in the shRN spacetime is given  by the Dirac equation,
\begin{equation}
\gamma^{\mu}D_{\mu}\Psi=0,
\end{equation} 
where $\Psi$ is the Dirac four-spinor, $\gamma^{\mu}$ are the coordinate dependent Dirac four-matrices and $D_{\mu}$ is spinor covariant derivative defined by,
\begin{equation}
D_{\mu}=\partial_{\mu}-\Gamma_{\mu}-i q A_{\mu}.
\end{equation}
Here $q$ is the charge of the Dirac field and $\Gamma_{\mu}$ are the spinor connection matrices. Following Refs.~\cite{huang_PRD_2017,dolan_cqg_2015} we decompose the Dirac four-spinor as 
\begin{equation}
\Psi=\frac{1}{\sqrt{r\sqrt{\Delta}}}\left(\begin{array}{c} -Q(r)S_1(\theta) \\ -P(r)S_2(\theta) \\  P(r)S_1(\theta) \\ Q(r)S_2(\theta) \end{array}\right) e^{i (m \phi-\omega t)}
\end{equation}
 and use the canonical orthonormal (symmetric) tetrad, proposed by Carter~\cite{carter_CMP_1968}, to yield two pairs of coupled first order differential equation,
\begin{eqnarray}\label{eq_radial_dirac1}
\sqrt{\Delta}\left(\frac{d}{dr}-\frac{i K}{\Delta}\right) P=\lambda Q,\label{eq_radial_dirac1}\\
\sqrt{\Delta}\left(\frac{d}{dr}+\frac{i K}{\Delta}\right) Q=\lambda P,\label{eq_radial_dirac2} 
\end{eqnarray}
and
\begin{eqnarray}
\left(\frac{d}{d \theta}+\frac{1}{2}\cot{\theta}-m \csc{\theta}\right)S_1=\lambda S_2,\\
\left(\frac{d}{d \theta}+\frac{1}{2}\cot{\theta}+m \csc{\theta}\right)S_2=\lambda S_1,
\end{eqnarray}	
where $K=\omega r^2-e q r$,  $-j\leq m\leq j$ and $\lambda=j+1/2$ $($ with $j=1/2,3/2...)$ is the separation constant.
The radial equations~(\ref{eq_radial_dirac1},\ref{eq_radial_dirac2}) can then be 
combined to yield,
\begin{equation}
\begin{split}
\sqrt{\Delta}&\frac{d}{dr}\left(\sqrt{\Delta}\frac{dP}{dr}\right)
\\&+\left(\frac{K^2+i (r-M)K}{\Delta}-2i \omega r+i e q-\lambda^2 \right)P=0.
\label{eq_combinedradial_dirac}
\end{split}
\end{equation}
If we define a new radial function, $\xi=\Delta^{-1/4}r P$, then equation~(\ref{eq_combinedradial_dirac}) can be written in a Schr\"{o}dinger like form in terms of the tortoise coordinate as
\begin{equation}
\frac{d^2\xi}{dr_{*}^2}+\tilde{W}(\omega,r)\xi=0,
\end{equation}
where
\begin{equation}
\label{eq_trts_dirac}
\begin{split}
\tilde{W}\left(\omega, r\right)= \frac{\Delta}{r^4}\left[\frac{\left(K+\frac{i}{2}\left(r-M\right)\right)^2}{\Delta}-2i \omega r+i e q\right.
\\\left.-\lambda^2-\frac{2M}{r}+\frac{2(e^2+s)}{r^2}\right].
\end{split}
\end{equation}
In the asymptotic limits of the tortoise coordinate equation~(\ref{eq_trts_dirac}) can be solved analytically with the QNM boundary conditions yielding
\begin{equation}
\label{eq_bc_dirac}
\xi \approx 
\begin{cases}
 e^{\frac{1}{4}\frac{\left(r_{+}-r_{-}\right)}{r_{+}^2}r_{*}-i \left(\omega-\frac{e q}{r_{+}} \right) r_*}&\mbox{ as \hspace*{2mm}}r_* \rightarrow -\infty \\
r_{*}^{\frac{1}{2}-i e q} e^{i\omega r_{*}}&\mbox{ as \hspace*{2mm}}r_* \rightarrow \infty.
\end{cases}
\end{equation} 
Equation~(\ref{eq_combinedradial_dirac}), similar to equation~(\ref{eq_radialKG}), also has two regular singularities at $r_{+}$ and $r_{-}$ and an irregular singularity as $r\rightarrow\infty$. So proceeding as before we introduce an ansatz, consistent with the boundary conditions~(\ref{eq_bc_dirac}),
\begin{equation}\label{eq_dirac_ansatz}
P=e^{i \omega r} (r-r_{-})^{\tilde{\rho}}\sum_{n=0}^{\infty}b_n u^{n+\tilde{\delta}},
\end{equation}
where  $u=\frac{r-r_{+}}{r-r_{-}}$,	 $\tilde{\rho}=-ieq+i\omega\left(r_{+}+r_{-}\right)$ and\\$\tilde{\delta} = \frac{1}{2}-\frac{i r_{+}^2 \left(\omega -\frac{e q}{r_{+}}\right)}{r_{+}-r_{-}}$.
Plugging (\ref{eq_dirac_ansatz})  back into equation~(\ref{eq_combinedradial_dirac}) we again arrive at the three term recurrence relations,
\begin{eqnarray}
\tilde{\alpha}_0 b_1 +\tilde{\beta}_0 b_0 & = & 0,\\
\tilde{\alpha}_n b_{n+1} +\tilde{\beta}_n b_n+\tilde{\gamma}_n b_{n-1} & = & 0,
\end{eqnarray}
\begin{figure*}
\centering
\includegraphics[width=1\linewidth]{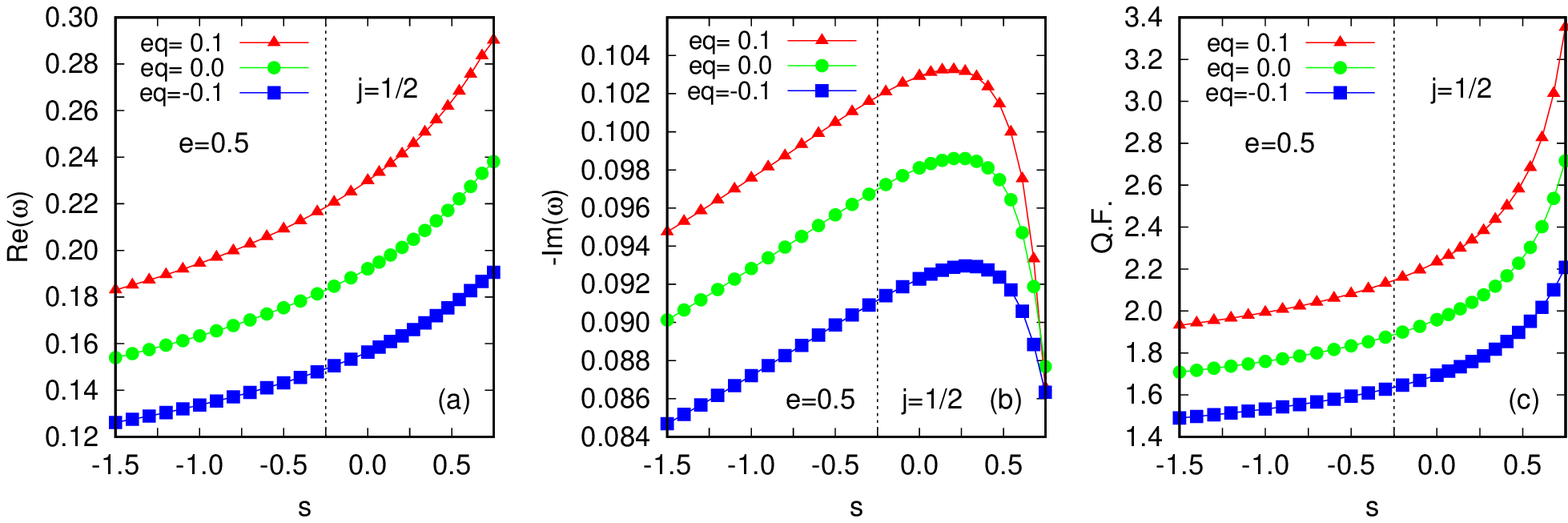}
\caption{Panels (a), (b) and (c), respectively, show the real and imaginary parts of the fundamental (Dirac) QN frequency and the quality factor as a function of $s$ for $j=1/2$ and $e=0.5$. Each curve in each panel corresponds to a particular value of $eq$ as indicated. The curve for $eq=0$ represents the perturbation by an uncharged Dirac field.  The vertical line denotes the value of $s$ (=-0.25) below which the spacetime behaves as mutated RN.}
\label{fig_7}
\vspace*{0.5cm}
\includegraphics[width=1\linewidth]{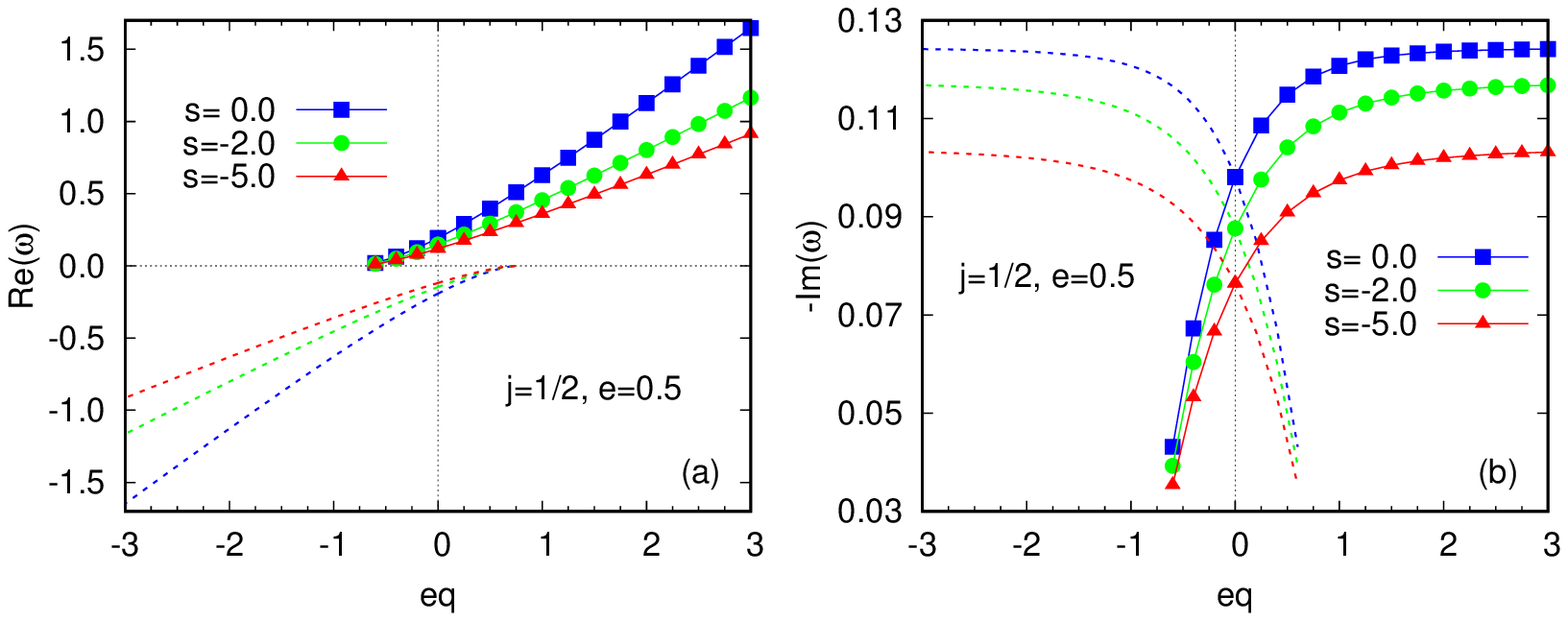}
\caption{Panels (a) and (b), respectively, show the real and imaginary parts of the fundamental (Dirac) QN frequency as a function of $eq$ for $j=1/2$ and $e=0.5$. Each curve in each panel corresponds to a particular value of $s$ as indicated. The dashed lines represent the symmetric curves due to the simultaneous transformation $eq\rightarrow -eq$ and $\omega\rightarrow -\omega^{*}$.}
\label{fig_8}
\end{figure*}
where
\begin{align}
\centering
\tilde{\alpha}_n&=(n+1) \left[\frac{1}{2} (2 n+3)+\frac{2 i r_{+} (e q-r_{+} \omega )}{r_{+}-r_{-}}\right],\\
\tilde{\beta}_n&=
\begin{aligned}[t]
&-\frac{r_{+}}{r_{+}-r_{-}} \bigg[-4 e^2 q^2-4 i r_{+} \omega  (3 i e q+2 n+1)\\
&\left.+6 i e n q+3 i e q+\lambda^2 +2 n^2+2 n-8 r_{+}^2 \omega ^2+\frac{1}{2}\right]\\
&+\frac{r_{-}\left[2 n (i e q+n+1)+i e q+\lambda^2 +\frac{1}{2}\right]}{r_{+}-r_{-}}\\
&-\frac{2 i (2 n+1) r_{-} r_{+} \omega}{r_{+}-r_{-}},
\end{aligned}\\[\jot]
\tilde{\gamma}_n&=
\begin{aligned}[t]
&-\frac{n+2 i (e q-\omega  (r_{-}+r_{+}))}{2 (r_{+}-r_{-})}[(2 n-1) r_{-}\\
&+r_{+} (-4 i e q-2 n+4 i r_{+} \omega +1)].
\end{aligned}
\end{align}
The convergence of the series~(\ref{eq_dirac_ansatz}) demands the recurrence  coefficients to satisfy an infinite continued fraction relation similar to equations~(\ref{eq_scalar_CF},\ref{eq_scalar_CF_inv}).\\
Applying Nollert's improvement, we now get the first three coefficients of the series~(\ref{eq_nollert_series}) as,\\
$C_0=-1$, $C_1=\sqrt{2i\omega\left(r_{-}-r_{+}\right)}$ and $C_2=\frac{5}{4}-i e q+2 i \omega r_{+}$.

\subsection{Numerical Results}
The behaviour of the real and imaginary parts of the fundamental QN frequency with the scalar charge is shown in Figs.~\hyperref[fig_7]{\ref*{fig_7}(a)} and \hyperref[fig_7]{(b)}. As before, we observe that for a fixed  value of the black hole electric charge, the magnitude of the imaginary part of the QN frequency for a given value of the field charge increases as the scalar charge changes from negative to positive and ultimately approaches the neutral one in the extremal limit. Thus, in the extremal limit the damping rate is independent of $s$. Here also we observe a peak in the magnitude of the imaginary part of the fundamental QN frequency for $s>-e^2$ whereas for $s<-e^2$, it decreases monotonically with the magnitude of the scalar charge. The real QN frequency on the other hand continues to grow with the scalar charge. Away from the extremal value of the scalar charge this causes the quality factor to grow steadily (see Fig.~\hyperref[fig_7]{\ref*{fig_7}(c)}), however as the extremal value of the scalar charge is reached the growth of the quality factor becomes quite rapid. 

Similar to the scalar case, we observe in Fig.~\ref{fig_8} that the QN frequency is symmetric with respect to the transformation $\left(eq\rightarrow -eq, \omega\rightarrow -\omega^{*}\right)$   and note that the critical value of electromagnetic interaction $(|eq|=0.7)$ at which the real part of the QN frequency vanishes, is almost unaffected by the presence of the scalar hair.

\section{Summary and Discussion}
\balance
In the present work we discussed the QN spectrum of massless and massive uncharged as well as charged scalar fields and massless charged Dirac fields in the vicinity of a charged spherically symmetric black hole with a scalar hair dubbed as the ``shRN'' black hole. We mainly focussed on negative values of the scalar charge with $s<-e^2$, for which the metric~(\ref{eq_metric}) represents  a ``mutated RN'' spacetime mimicking the Einstein-Rosen bridge. 

Unlike the appearance of a distinct peak in the magnitude of the imaginary part of the fundamental QN frequency of an shRN black hole for scalar and Dirac fields with $s>-e^2$, the mutated RN spacetime $(s<-e^2)$ is characterised by monotonically decreasing $|Im(\omega)|$ (see Figs.~\hyperref[fig_2]{\ref*{fig_2}(b)}, \hyperref[fig_5]{\ref*{fig_5}(f)} and \hyperref[fig_7]{\ref*{fig_7}(b)}). For uncharged fields, the shRN black hole effectively behaves as an RN black hole with effective electric charge, $e_{eff}=\sqrt{e^2+s}$, provided $s$ lies in $M^2-e^2\geq s> -e^2$.

For massive scalar field, the phenomenon of \textit{quasi-\\resonance}, characterised by vanishingly small $|Im(\omega)|$ is observed. We also showed the behaviour of the quality factor with the scalar charge for both the scalar and Dirac fields. As the extremal limit is approached either by increasing the electric charge for a fixed $s$ or vice-versa, we find that the imaginary  part of $\omega$ for neutral and charged scalar or Dirac perturbations to be coincident. 

In the presence of electric charge of the perturbing fields, we observe the existence of a critical value of $|eq|$, above which the real part of the QN frequency vanishes, for both the scalar and Dirac fields. This value is completely unaffected by the presence of  scalar hair.

Following the method of Cho~\cite{cho_PRD_2006}, we start with the asymptotic form of the Dirac QN frequency and calculated the area spectrum of the shRN black hole based on the proposals of Kunstatter~\cite{kunstatter_PRL_2003} and Maggiore~\cite{maggiore_PRL_2008}. We obtain the area quantum as $\Delta A=8 \pi \hbar$. This being the same as that of an RN black hole~\cite{ortega_CQG_2011}, we refrain from including a detailed calculation of the same. 

Very recently Saleh, Thomas and Kofane~\cite{saleh_ASS_2014,saleh_ASS_2016} discussed the QN spectrum of massless uncharged scalar and Dirac fields in the vicinity of a ``quantum-corrected'' Schwarz\-schild black hole~\cite{kim_PLB_2012} using 3rd order WKB approximation. The metric used by them is effectively similar to that of the mutated RN spacetime discussed in the present work. The results obtained by us for the massless uncharged scalar and Dirac fields in the shRN background (with $s<-e^2$) using the more accurate continued fraction method, is qualitatively similar to them. The present work is, however, much more general as it includes charge for both the scalar and the Dirac fields and mass for the scalar field.

The QN spectrum analysis was also carried out with the 3rd order WKB approximation which is generally believed to be less accurate. We add a table in the appendix comparing the results for one example, namely that for an uncharged massless scalar perturbation of the shRN black hole. The WKB approximation is known to yield more and more accurate results for higher and higher values of multipole ($l$). The table contains the values given by the Leaver method and that by the WKB approximation for the real and the imaginary parts of the quasinormal mode frequencies for $l=1, 2, 3$. It is apparent from the table that the difference in the results given by the two methods reduces for higher values of $l$.

We point out that in the present work we have not observed any QNM with positive imaginary part indicating the stability of the shRN black hole under  massive (and massless)   charged (and uncharged)  scalar perturbations as well as under massless charged (and uncharged) Dirac perturbations for both $s$ in $[-e^2,M^2-e^2)$ and in the ``mutated'' regime,  $s<-e^2$. This implies that the mutated RN spacetime is also stable under all the above mentioned types of perturbation.

It has already been pointed out in section~\ref{sec_2} that the scalar field in this case is a constant ($\psi = \pm\sqrt{\frac{6}{8 \pi G}}\sqrt{\frac{s}{s+e^2}}$). From equation~(\ref{eq_action}), it is clear that the action can be transformed into an Einstein-Maxwell system, with a different value for the effective Newtonian constant of gravity $G_{eff} = \frac{G}{(1 - \frac{4\pi G}{3}){\psi}^2}$. However, the spacetime is qualitatively different if $s$ is negative, and $|s| > e^2$. The resulting metric component is\\ $\left(1-\frac{2M}{r}-\frac{e_{eff}^2}{r^2} \right)$ as opposed to $\left(1-\frac{2M}{r}+\frac{e_{eff}^2}{r^2} \right)$ in the standard RN spacetime. The distinctive features of this metric has already been discussed at length.

Finally, we mention that the qualitative features for the quasinormal modes for the charged or uncharged massive scalar fields and also that for the charged Dirac field for a mutated RN (shRN with $s<-e^2$) background are qualitatively same as that for a usual RN black hole except for the complete monotonic behaviour of the damping (imaginary part of the QNM) in the case of  the former as opposed to the existence of a peak in the latter case, an RN black hole.

\begin{acknowledgements}
NB wishes to thank Sayan Kar for drawing his interest to the mutated RN metric.
\end{acknowledgements}

\appendix
\onecolumn
\section*{Appendix}\label{app}
\begin{table}[!htb]
\caption{Fundamental QN frequencies of massless uncharged scalar fields in the background of an shRN black hole of mass, $M=1$ and electric charge, $e=0$ for different values of the multipole index and scalar charge. For each value of $s$, the first line is obtained using the continued fraction method with $175$ terms and the second line is obtained using the 3rd order WKB approximation.}
\begin{tabular*}{\textwidth}{c @{\extracolsep{\fill}} rrrrrr} 
\hline\hline
\multicolumn{3}{c}{\hspace{0.065\textwidth}$l=1$}                                                                                                         & \multicolumn{2}{c}{$l=2$}                                                                           & \multicolumn{2}{c}{$l=3$}                                                                            \\ 
\hline
\multicolumn{1}{l}{$s$}   & \multicolumn{1}{l}{$Re(\omega)$} & \multicolumn{1}{l}{$Im(\omega)$} & \multicolumn{1}{l}{$Re(\omega)$} & \multicolumn{1}{l}{$Im(\omega)$} & \multicolumn{1}{l}{$Re(\omega)$} & \multicolumn{1}{l}{$Im(\omega)$}  \\ 
\hline
\multirow{2}{*}{0.99}       & 0.3762055912                                      & -0.0900896009                                     & 0.6237563062                                      & -0.0895373909                                     & 0.8720041949                                      & -0.0893846794                                      \\
                            & 0.3742670229                                      & -0.0900696201                                     & 0.6232727045                                      & -0.0895182148                                     & 0.8718206816                                      & -0.0893779664                                      \\
\multirow{2}{*}{0.9}        & 0.3637066222                                      & -0.0947427505                                     & 0.6009488609                                      & -0.0942904001                                     & 0.8393317935                                      & -0.0941634988                                      \\
                            & 0.3618775351                                      & -0.0947062218                                     & 0.6004948731                                      & -0.0942713464                                     & 0.8391598515                                      & -0.0941570756                                      \\
\multirow{2}{*}{0.7}        & 0.3408752613                                      & -0.0986496604                                     & 0.5624223918                                      & -0.0979828596                                     & 0.7852194051                                      & -0.0977944238                                      \\
                            & 0.3393237655                                      & -0.0987271148                                     & 0.5620307364                                      & -0.0979828452                                     & 0.7850707716                                      & -0.0977934267                                      \\
\multirow{2}{*}{0.5}        & 0.3235342925                                      & -0.0993515868                                     & 0.533818086                                       & -0.0985724807                                     & 0.7452888813                                      & -0.0983503413                                      \\
                            & 0.3219858221                                      & -0.0995455106                                     & 0.5334311042                                      & -0.0985914448                                     & 0.7451428561                                      & -0.098354732                                       \\
\multirow{2}{*}{0.3}        & 0.3096415907                                      & -0.0989749919                                     & 0.5110162987                                      & -0.0981311196                                     & 0.7135039562                                      & -0.0978893749                                      \\
                            & 0.307999452                                       & -0.0992457629                                     & 0.5106143307                                      & -0.0981635738                                     & 0.7133535414                                      & -0.0978977124                                      \\
\multirow{2}{*}{0.1}        & 0.2980681332                                      & -0.0981534347                                     & 0.4920504262                                      & -0.0972678241                                     & 0.6870780311                                      & -0.0970134153                                      \\
                            & 0.2963068621                                      & -0.0984754655                                     & 0.491628113                                       & -0.097310014                                      & 0.6869212869                                      & -0.0970246966                                      \\
\multirow{2}{*}{0}          & 0.2929361333                                      & -0.0976599889                                     & 0.4836438722                                      & -0.096758776                                      & 0.6753662325                                      & -0.0964996277                                      \\
                            & 0.2911141164                                      & -0.0980013631                                     & 0.4832110304                                      & -0.0968048549                                     & 0.675206178                                       & -0.0965121143                                      \\
\multirow{2}{*}{-0.1}       & 0.2881615316                                      & -0.0971353427                                     & 0.4758233999                                      & -0.0962210405                                     & 0.6644712038                                      & -0.0959579129                                      \\
                            & 0.2862799244                                      & -0.0974930215                                     & 0.4753801758                                      & -0.0962705126                                     & 0.6643078766                                      & -0.0959714677                                      \\
\multirow{2}{*}{-0.3}       & 0.2795122233                                      & -0.0960340955                                     & 0.4616560916                                      & -0.0950993038                                     & 0.6447339818                                      & -0.0948299372                                      \\
                            & 0.2775174544                                      & -0.0964175448                                     & 0.4611930519                                      & -0.0951543929                                     & 0.6445644025                                      & -0.0948453003                                      \\
\multirow{2}{*}{-0.5}       & 0.2718459161                                      & -0.0949062659                                     & 0.4490967969                                      & -0.0939565158                                     & 0.6272360885                                      & -0.0936825807                                      \\
                            & 0.2697477256                                      & -0.0953089565                                     & 0.4486155668                                      & -0.0940160429                                     & 0.6270607761                                      & -0.093699416                                       \\
                            & 0.2661727046                                      & -0.0947528304                                     & 0.442827786                                       & -0.0934480225                                     & 0.6190057789                                      & -0.093128333                                       \\
\multirow{2}{*}{-0.7}       & 0.2649696261                                      & -0.0937815642                                     & 0.4378289057                                      & -0.0928207752                                     & 0.6115362336                                      & -0.0925434544                                      \\
                            & 0.2627780753                                      & -0.0941990015                                     & 0.4373312051                                      & -0.092883877                                      & 0.6113557438                                      & -0.0925615095                                      \\
\multirow{2}{*}{-0.9}       & 0.2587422561                                      & -0.0926759376                                     & 0.4276214399                                      & -0.0917069907                                     & 0.5973126872                                      & -0.0914271526                                      \\
                            & 0.2564668676                                      & -0.0931049246                                     & 0.4271089136                                      & -0.0917730128                                     & 0.597127553                                       & -0.0914462318                                      \\
\multirow{2}{*}{-1}         & 0.2558376364                                      & -0.0921331191                                     & 0.4228593181                                      & -0.0911609447                                     & 0.5906764879                                      & -0.0908801044                                      \\
                            & 0.2535236413                                      & -0.0925669597                                     & 0.4223399545                                      & -0.0912282303                                     & 0.5904892182                                      & -0.0908996369                                      \\
\multirow{2}{*}{-1.1}       & 0.2530574278                                      & -0.0915978709                                     & 0.4183004849                                      & -0.0906229436                                     & 0.5843232974                                      & -0.0903412442                                      \\
                            & 0.2507068851                                      & -0.0920360624                                     & 0.4177746428                                      & -0.0906913813                                     & 0.5841340084                                      & -0.0903611949                                      \\
\hline\hline
\end{tabular*}
\end{table}
\twocolumn

\end{document}